\magnification\magstephalf

\tolerance 10000
\settabs 6\columns

\input epsf.tex
\font\rfont=cmr10 at 10 true pt
\def\ref#1{$^{\hbox{\rfont {[#1]}}}$}
\def\P{{I\!\!P}}
\def\hat{\bar}


\font\fourteenbf=cmbx12 scaled\magstep1


\def\pd {\partial}
\def\pmb#1{\setbox0=\hbox{#1}
 \kern.05em\copy0\kern-\wd0 \kern-.025em\raise.0433em\box0 }

\def \half {{\scriptstyle {1 \over 2}}}

\def \quarter {{\scriptstyle {1 \over 4}}}

 %


\def\boxit#1{\vbox{\hrule\hbox{\vrule\kern1pt\vbox
{\kern1pt#1\kern1pt}\kern1pt\vrule}\hrule}}

\def \cl {\centerline}
\def\h{\hfill\break}
\parskip=6pt
\parindent=0pt
\hsize=17truecm\hoffset=-5truemm
\voffset=5truemm\vsize=24.5truecm
\def\footnoterule{\kern-3pt
\hrule width 17truecm \kern 2.6pt}


\catcode`\@=11 

\def\nolabels{\def\wrlabeL##1{}\def\eqlabeL##1{}\def\reflabeL##1{}}
\def\writelabels{\def\wrlabeL##1{\leavevmode\vadjust{\rlap{\smash%
{\line{{\escapechar=` \hfill\rlap{\sevenrm\hskip.03in\string##1}}}}}}}%
\def\eqlabeL##1{{\escapechar-1\rlap{\sevenrm\hskip.05in\string##1}}}%
\def\reflabeL##1{\noexpand\llap{\noexpand\sevenrm\string\string\string##1}}}
\nolabels
\global\newcount\refno \global\refno=1
\newwrite\rfile
\def\defref{$^{{\hbox{\rfont [\the\refno]}}}$\nref}
\def\nref#1{\xdef#1{\the\refno}\writedef{#1\leftbracket#1}%
\ifnum\refno=1\immediate\openout\rfile=refs.tmp\fi
\global\advance\refno by1\chardef\wfile=\rfile\immediate
\write\rfile{\noexpand\item{#1\ }\reflabeL{#1\hskip.31in}\pctsign}\findarg}
\def\findarg#1#{\begingroup\obeylines\newlinechar=`\^^M\pass@rg}
{\obeylines\gdef\pass@rg#1{\writ@line\relax #1^^M\hbox{}^^M}%
\gdef\writ@line#1^^M{\expandafter\toks0\expandafter{\striprel@x #1}%
\edef\next{\the\toks0}\ifx\next\em@rk\let\next=\endgroup\else\ifx\next\empty%
\else\immediate\write\wfile{\the\toks0}\fi\let\next=\writ@line\fi\next\relax}}
\def\striprel@x#1{} \def\em@rk{\hbox{}} 
\def\lref{\begingroup\obeylines\lr@f}
\def\lr@f#1#2{\gdef#1{\defref#1{#2}}\endgroup\unskip}
\newwrite\lfile
{\escapechar-1\xdef\pctsign{\string\%}\xdef\leftbracket{\string\{}
\xdef\rightbracket{\string\}}}

\def\writestop{\def\writestoppt{\immediate\write\lfile{\string\p
ageno%
\the\pageno\string\startrefs\leftbracket\the\refno\rightbracket%
\string\def\string\secsym\leftbracket\secsym\rightbracket%
\string\secno\the\secno\string\meqno\the\meqno}\immediate\closeout\lfile}}
\def\writestoppt{}\def\writedef#1{}
\catcode`\@=12 

\rightline{ULG-PNT-96-1-JRC}
\rightline{M/C-TH 96/02}
\rightline{DAMTP-96/02}
\bigskip
\centerline{\fourteenbf THE BFKL POMERON: CAN IT BE DETECTED?}
\bigskip
\centerline{J R Cudell}
\centerline{Institut de Physique, Universit\'e de Li\`ege}
\vskip 5pt
\centerline{A Donnachie}
\centerline{Department of Physics, Manchester University}
\vskip 5pt
\centerline{P V Landshoff}
\centerline{DAMTP, Cambridge University$^*$}
\footnote{}{$^*$ email addresses: cudell@gw.unipc.ulg.ac.be,
\ ad@a3.ph.man.ac.uk, \ pvl@damtp.cam.ac.uk}
\bigskip
{\bf Abstract}

An estimate is derived for the absolute magnitude of 
BFKL pomeron exchange at $t=0$. 
The analysis takes account of energy conservation and of the need realistically
to model nonperturbative contributions to the BFKL integral from
infrared regions. Experiment finds that there is little or no room for a
significant BFKL term in soft processes, and this constrains its magnitude in
hard and semihard processes, so that it is unlikely to be detectable.

\vskip 8 truemm
{\bf 1 Introduction}

Total cross-sections for hadron-hadron and photon-hadron collisions all
seem to increase\defref\total{
A Donnachie and P V Landshoff, Physics Letters B296 (1992) 227
} 
at high energy as the same very-slowly varying power of the
energy, $s^{0.08}$. This is said to be characteristic of
soft pomeron exchange and is an inherently nonperturbative phenomenon.
While experiment finds\defref\diff{
H1 collaboration: T Ahmed et al, Physics Letters B348 (1995) 681\h
ZEUS collaboration: M Derrick et al, Physics Letters B356 (1995) 129
} 
that the soft pomeron is exchanged
also in diffractive electroproduction at quite high
$Q^2$, there are other high-$Q^2$ 
data in which a somewhat more rapid variation with
energy is found. These are data for the small-$x$ behavior of $\nu W_2$,
which seems\defref\smallx{
H1 collaboration: T Ahmed et al, Nuclear Physics B439 (1995) 471\h
ZEUS collaboration: M Derrick et al, Z Phyzik C65 (1995) 379
} 
to be more like $f(Q^2)(W^2)^{0.3}$ than 
$f(Q^2)(W^2)^{0.08}$, and exclusive $J/\psi$ photoproduction 
and $\rho$ electroproduction, which again seem
to behave\defref\rise{
ZEUS collaboration: M Derrick et al, Physics Letters B356 (1995) 601
}
more like $[f(Q^2)(W^2)^{0.3}]^2$. It is not yet clear what is the
cause of this more violent variation with energy. 

A candidate explanation
is that the perturbative BFKL pomeron is responsible\defref\bfkl{
E A Kuraev, L N Lipatov and V Fadin, Soviet Physics JETP 45 (1977) 199\h
Y Y Balitskii and L N Lipatov, Sov J Nuclear Physics 28 (1978) 822
}. 
In this paper we argue that this explanation is unlikely
to be correct: while the power of $W$ predicted by the BFKL equation
can fit the observed behaviour, the magnitude of the constant that
multiplies it is almost certainly much too small.

A clean calculation of this magnitude is not possible, because 
one cannot cleanly separate the perturbative and nonperturbative effects.
This problem arises already in lowest order.
The crudest model\defref\low{
F E Low, Physical Review D12 (1975) 163\h
S Nussinov, Physical Review Letters 34 (1975) 1286
} 
for pomeron exchange is simple two-gluon exchange between quarks,
figure 1. At large $s$ this gives a constant cross-section:
$$\sigma _0= {8\over 9}\int d^2k_T\;{\alpha _s^2\over k_T^4}
\eqno(1.1)
$$
Here $k_T$ is the transverse momentum of the internal quark lines, which
correspond to final-state  jets.
It is unsafe to use this perturbative formula for the production of quark
jets with too low a transverse momentum,
$k_T^2 < 1$ GeV$^2$, say. This is because the squared gluon 4-momentum
asymptotically is just $k^2\sim -k_T^2$, and it is illegal to
extend the integration into the nonperturbative region.
If we exclude this nonperturbative
region from the integration in (1.1), then for fixed $\alpha _s$
we obtain a quark-quark cross-section
equal to $1.1\,\alpha_s ^2$ mb, which for any reasonable perturbative value of
$\alpha _s$ is considerably less than the observed cross-section of a few mb: 
if the lowest-order calculation is a good guide
most of the cross-section comes from the nonperturbative region\defref\otto{
P V Landshoff and O Nachtmann, Z Physik C35 (1987) 405
}. 
We argue in this paper that a similar result holds when we include
higher orders, and that it applies not only to purely soft processes
like total hadron-hadron cross sections, but also to semihard ones such
as exclusive vector production or the small-$x$ behaviour of structure
functions.
\topinsert
\centerline{{\epsfxsize=30truemm\epsfbox{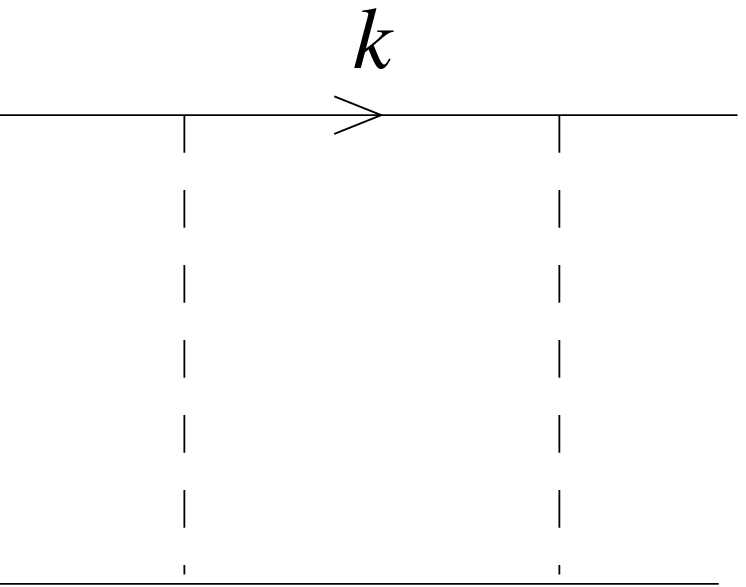}}}\hfill\break
\centerline{Figure 1: Exchange of two gluons between a pair of quarks.}
\endinsert

We begin in the next section by exploring further
the exchange of the perturbative pomeron between a pair of
quarks at $t=0$. This is generated by the BFKL equation.
In its simplest form, the BFKL equation 
describes asymptotically large energies, where energy conservation
constraints have become unimportant.
Previous attempts\defref\cl{
J C Collins and P V Landshoff, Physics Letters B276 (1992) 196
}\defref\forshaw{
M F McDermott, J R Forshaw and G G Ross, Physics Letters B349 (1995) 1895\h
J Forshaw, P N Harriman and P Sutton, Nuclear Physics B416 (1994) 739
}
to impose energy conservation have been unsatisfactory in two
respects. The BFKL equation
takes an input amplitude Im $T_0(s)$ and modifies it by  real and
virtual gluon corrections. These two types of correction need to be handled
differently.
Energy conservation restricts the {\it sum} of the
transverse energies of all the real gluons to be less than $\surd s$,
while previous work has imposed this constraint on
just their individual energies and has applied it also to the virtual
gluons, which is not correct.

Energy conservation imposes a cut-off at the high-momentum end of the
loop integrations in the BFKL equation. As we have already indicated,
the low-momentum end also needs 
attention, since the BFKL equation works with perturbative gluon propagators.
Because of confinement effects, at small $k^2$ the gluon propagator
receives very significant nonperturbative corrections\defref\cr{
D Zwanziger, Nuclear Physics B323 (1989) 513\h
U Habel, R Konning, H-G Reusch, M Stingl and S Wigard, Z Physik A336 (1990) 423, 435\h
J R Cudell and D A Ross, Nucl Phys B359 (1991) 247
} 
so that, even if the BFKL equation has a finite solution with a purely
perturbative propagator, this solution makes no physical sense. There have been
several attempts to take this into account, none of them very 
satisfactory\ref{\cl}\ref{\forshaw}\defref\ross{
R E Hancock and D A Ross, Nuclear Physics B383 (1992) 575\h
A J Askew et al, Physical Review D49 (1994) 4402
}. 
They either simply exclude the low-$k^2$ part of the loop integration,
or they try to use a nonperturbative propagator at low $k^2$,
or they use a nonperturbative input
amplitude $T_0$,  which can be only part of the solution. 
In this paper we attempt to improve on this, though inevitably we cannot
deal with two issues that arise: that of gauge invariance, and
whether the BFKL equation itself, and not just the gluon propagator,
must not also be modified.

In section 3, we initially  impose a lower cut-off $\mu$ on the transverse
momenta of the real gluons. That is, at first we calculate only a small part 
$\sigma{(K_T>\mu)}$ of
the total cross-section for quark-quark scattering, arising from events 
where the final state consists
only of any number of partons of
transverse momentum greater than $\mu$.
The question arises: what is the minimum choice for $\mu$ such that
the perturbative calculation of $\sigma{(K_T>\mu)}$ is likely to
be trustworthy? By comparing our calculation with $pp$ and $\bar pp$
total cross section data, we show that it is unsafe to take $\mu$
to be less than 2 GeV.

In section 4 we
take account of the fact that it is extremely rare that {\it all} the
partons will have ${K_T>\mu}$.  
In a general event, we may group  the final-state partons according to
their rapidities. As there is no transverse-momentum ordering, their
transverse momentum is not correlated with their rapidity. So as we
scan the rapidity range we find groups of partons all having transverse
momentum greater than $\mu$, with each such group separated by a group
in which none of the partons has transverse momentum greater than ${\mu}$.
This we show in figure 2a, where the heavy lines have transverse momentum
${K_T>\mu}$, while the light lines have ${K_T<\mu}$.
When we sum over all possible numbers of lines in a group with
${K_T>\mu}$ we obtain the hard pomeron $\P _H$ which we calculate
in section 3, while a group with ${K_T<\mu}$ sums to a soft exchange
$\P _S$. So the result is figure 2b.
When we sum over all final states, we obtain terms
$$
\P _S \;+\;\P _H \;+\; \P _S\otimes\P _H \;+\; \P _H\otimes\P _S
\;+\;\P _S\otimes\P _H\otimes\P _S \;+\;\dots
\eqno(1.2)
$$
The separate terms here each depend on the value chosen for $\mu$, but of
course the sum must not. As we explain in section 4, this means that
$\P _S$ is not exactly the contribution from soft pomeron exchange, only
nearly so when $\mu$ is large enough.
In section 4 we analyse whether
we may expect to obtain an enhancement of the fiercely-varying part
of the cross-section by including in this way
also final states where only a subset
of the partons have transverse momentum greater than $\mu$. That is, we
ask whether mixing in contributions from soft interactions can
very significantly  enhance
the magnitude of the contribution from the hard ones. Our conclusion is
that such enhancement is at the very most an order of magnitude.
\topinsert
\centerline{{\epsfxsize=90truemm\epsfbox{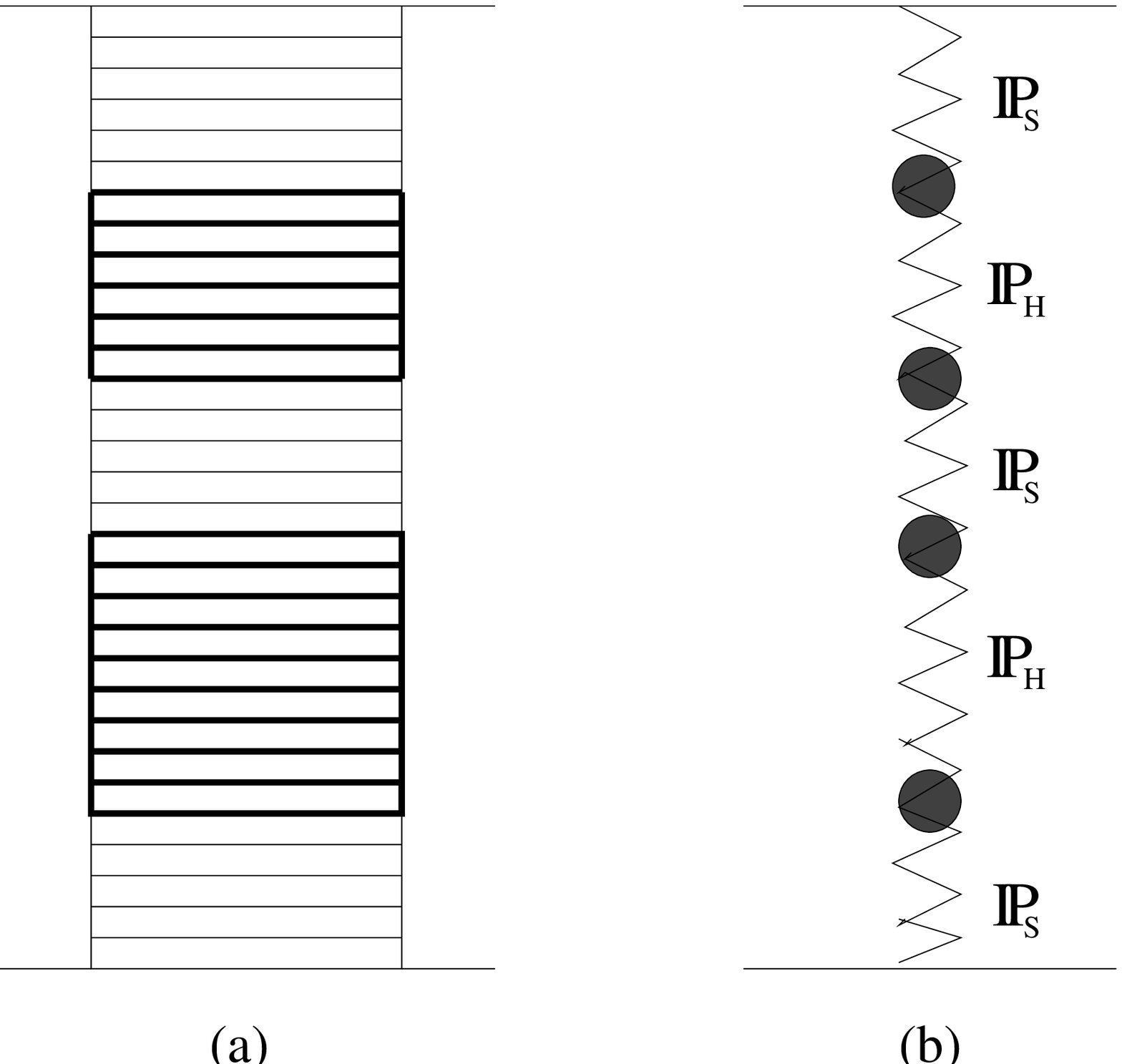}}}\hfill\break
Figure 2: (a) alternating groups of partons with low and high $K_T$,
with (b) their sum giving alternating soft and hard pomerons.
\endinsert

In section 5 we take a first look at a semisoft process, using
$\gamma ^*p\to\rho p$ as an example.
While we do not expect that  a ``soft'' process such as the total cross-section
for quark-quark scattering should receive most of its contribution from states
containing only partons having transverse momentum greater than 2 GeV,
for semihard processes things might be different\defref\bartels{
J Bartels, H Lotter and M Vogt, DESY-95-224, hep-ph/9511399
}. 
We find that, although as $Q^2$ increases it is true that high-transverse-%
momentum partons become relatively more likely,
the hard-pomeron-exchange
contribution is tiny at HERA energies. 
Section 6 is devoted to a first look at a purely hard process,
$\gamma ^* \gamma ^*\to \rho\rho$ at large $Q^2$. Again we find that the
perturbative contribution to the amplitude is tiny, but we verify that
in such a hard process the ``diffusion'' ideas\ref{\bartels} about
the magnitude of the parton transverse momenta are valid.

Finally, in section 7 we discuss various points. Our discussion throughout
is confined to $t=0$, where hard-pomeron exchange has to compete with
the large contribution from soft exchange.

\bigskip
{\bf 2 The BFKL cross-section}

We consider purely-gluon exchange between a pair of zero-mass
quarks at $t=0$ in  
3-colour QCD. We begin by recapitulating the calculation of the lowest-order
graph, figure 1. Change integration variables from $k$ to $(x,y,k_T)$,
where
$$
k=xp+yp'+k_T
\eqno(2.1)
$$
We need the imaginary part of the amplitude, for which the two internal
quarks $K_1$ and $K_2$ are on shell. The 
$\delta$-functions that put them on shell
give
$$x=-y=\half (-1+R)$$$$
R=\sqrt{1-4k_T^2/s}
\eqno(2.2)
$$
Because, for large $\surd s$, the main contribution to the integral comes from
values of $k_T$ much less than $\surd s$, we may approximate 
$$
x=-y\approx k_T^2/s$$$$
k^2=xys-k_T^2\approx -k_T^2
\eqno(2.3)
$$
These approximations then give the cross-section
$$\sigma _0= \int {d^2k_T\over k_T^4}\;t_0(k_T^2)$$$$
t_0(k_T^2)={8\alpha _s^2\over 9}
\eqno(2.4)
$$
Because we want to
calculate the cross-section for production of a pair of partons each having
transverse momentum greater than some fixed value $\mu$, 
we confine the integration
to $|k_T^2|>\mu ^2$. This will also prevent the integration extending
into the region where the two propagators that carry momentum 
$k$ are nonperturbative. 
In any case,  we
cannot simply integrate (2.4) down to $k_T=0$; not only would this give an
infrared divergence, it would also not provide any means of giving the 
cross-section its correct dimension%
\footnote{a}{
Other authors\defref\gs{
J F Gunion and D Soper, Physical Review D15 (1977) 2617
} introduce a dimension into hadron-hadron scattering (though not
quark-quark scattering) through a form factor,
which is another way of bringing in a nonperturbative effect. In the 
following, we shall cut off the transverse momentum of the 
real emission at a scale $\mu>1
 $ GeV $>>1/1$ fm. This means that the quark momenta entering the form 
factor will be very
 different, unless the gluons are coupled to the same quark. Hence the extra 
terms which lead to an infrared stable answer for $\mu=0$, and which involve 
different 
quarks within the hadrons, are negligible in the cut-off case. This is of 
course not true at high-$Q^2>\mu^2$, and we shall take them into 
account when we consider semihard and hard processes.}.

It is evident from (2.2) that there must also be an upper limit, in order
that $R$ be real: 
$$
2k_T<\surd s
\eqno(2.5)
$$
This is just the condition that the total transverse energy
of the real partons is no more than $\surd s$.
Of course, an
exact calculation would not merely impose this kinematic constraint on the
asymptotic form of the integrand associated with a given graph; it would
also include nonasymptotic terms in the integrand. However, this
is almost impossible to achieve beyond the lowest order, 
and so we shall be content with 
simply imposing the kinematic constraints.
When $\surd s\gg\mu$ the upper limit (2.5) has little effect, but when 
we consider
the production of a large number of 
partons, it becomes important. The
study of this is one subject of our paper.

Cheng and Wu\defref\chengwu{
H Cheng and T T Wu, {\it Expanding protons}, MIT Press (1987)
}
have calculated the sum of the order $\alpha _s^3$ graphs. Their result may
be written in a form that makes contact with the BFKL equation, as follows. 
Write the imaginary part of the $\alpha _s^{n+2}$ contribution to the 
cross-section as
$$
\sigma _n(s)={(\log\hat s)^n\over n!}\int {d^2k_n\over k_n^4}\;t_n(k_n)$$$$
\hat s ={s\over\mu ^2}
\eqno(2.6)
$$
It is just a guess that the appropriate scale for $s$ here is $\mu ^2$; to
check this would require an almost-impossible nonleading calculation.
But it does seem the most reasonable guess.
Then the imaginary part of the sum of the order $\alpha _s^3$ graphs
at high $s$ may be written in the form 
$$
t_1(k_1)= K\otimes t_0
\eqno(2.7a)
$$
where
$$
K\otimes t_0=\int d^2k_0\; K(k_1,k_0)\left\{t_0(k_0)-\half t_0(k_1)\right\}
\eqno(2.7b)
$$
with\ref{\chengwu}
$$
K(k_1,k_0)= {3\over\pi ^2} {\alpha_s\; k_1^2\over k_0^2(k_0-k_1)^2}
\eqno(2.7c)
$$
Written in this way,
the relation (2.7) between the $n=1$ and $n=0$ terms is just the BFKL relation:
The first term in the curly bracket is associated with real gluons, and the
second with virtual. The virtual-gluon term may be written in a more
familiar form\defref\cn{
J R Cudell and B U Nguyen, Nuclear Physics B420 (1994) 669
},
by using the simple identity
$$
{1\over k_0^2(k_0-k_1)^2}={1\over k_0^2[k_0^2+(k_0-k_1)^2]}
+{1\over (k_0-k_1)^2[k_0^2+(k_0-k_1)^2]}
$$
When we subject this to the integration in (2.7), the last two terms
contribute equally, as can be seen by the change of integration variable
$$
k_0\to k_1-k_0
$$
Hence we can write (2.7) in a form that has become more familiar\ref{\cl}
$$
t_1(k_1)
={3\over\pi ^2}\int d^2k_0{\alpha_s\; k_1^2\over(k_1-k_0)^2}\left\{
{t_0(k_0)\over k_0^2}-{t_0(k_1)\over k_0^2
+(k_0-k_1)^2}
\right\}
\eqno(2.8)
$$

We shall
assume that (2.7) generalises to higher values of $n$:
$$
t_n(k_n)=\int d^2k_{n-1}K(k_n,k_{n-1})
\left\{t_{n-1}(k_{n-1})-\half t_{n-1}(k_n)\right\}
\eqno(2.9a)
$$
with $K$ given in (2.7c), so that
$$
t_n=K\otimes K\otimes \dots \otimes t_0
\eqno(2.9b)
$$
Then, in the absence of any cut-offs on the $k$ integrations
and with fixed coupling $\alpha _s$,
the infinite sum of $\sigma _n$
over $n$ yields\ref{\bfkl ,\cl} just the familiar power of $s$
$$
{12 \alpha _s\over\pi}\log 2
\eqno(2.10)
$$

\bigskip
{\bf 3 The cut-off BFKL equation}

We may write (2.9a) as
$$
t_n(k_n)=\int d^2k_{n-1}K(k_n,k_{n-1})t_{n-1}(k_{n-1})\;-\;\half
\phi (k_n^2)t_{n-1}(k_n)
\eqno(3.1)
$$
where
$$
\phi (k^2)={3\over\pi ^2}\int d^2q{\alpha _s\; k^2\over q^2(k-q)^2}
\eqno(3.2a)
$$
In the absence of an infrared cut-off, each term in (3.1) is separately
divergent, but the divergences cancel between them. As we have explained,
the fact that this cancellation of divergences occurs does not imply
that it is meaningful, because without an infrared cut-off the integration
extends illegally into the nonperturbative domain. 
\topinsert
\centerline{{\epsfxsize=50truemm\epsfbox{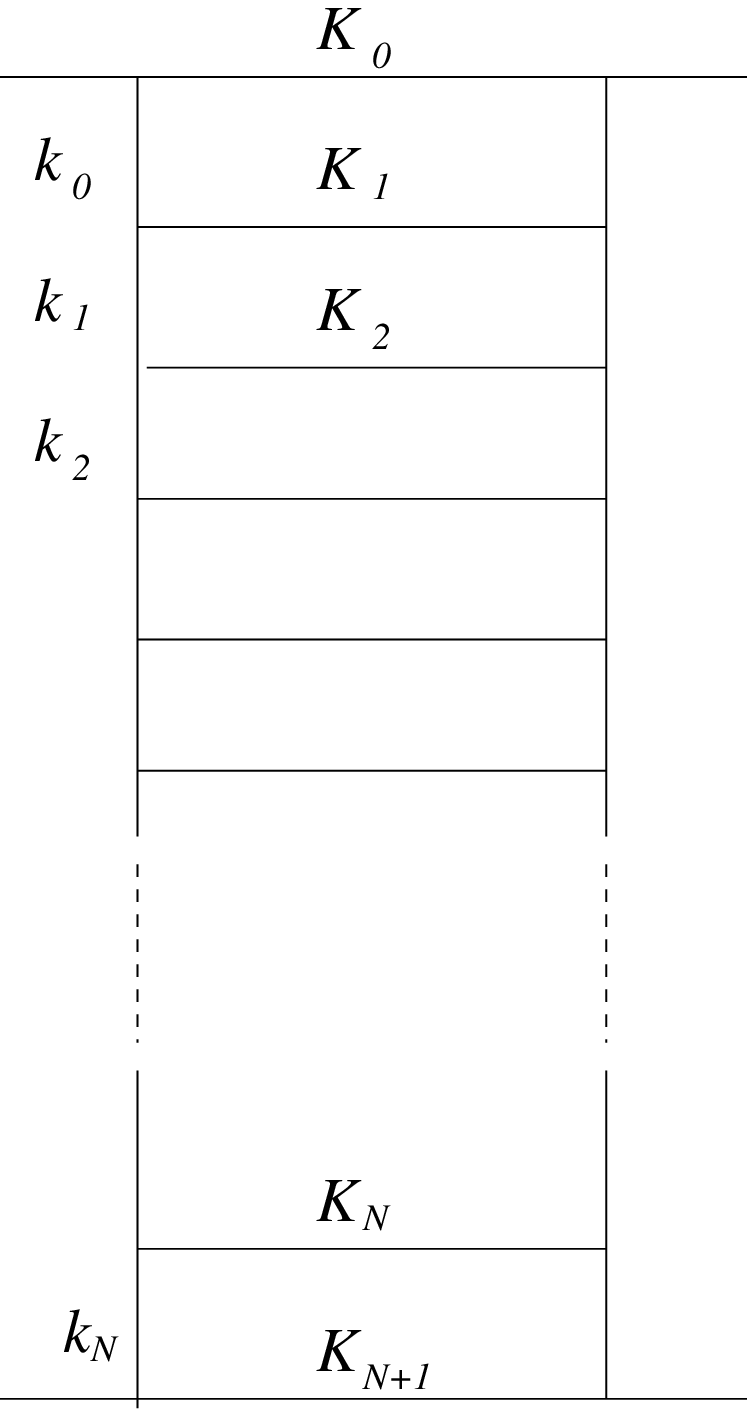}}}\hfill\break
\centerline{Figure 3: Kinematics of the BFKL ladder}
\endinsert

The function $\phi$ represents the virtual-gluon insertions. When the
divergence has been regulated somehow, we may resum these 
insertions. In order to do this, first write
$$
{(\log\hat s)^n\over n!}={1\over 2\pi i}\int {dj \over j^{n+1}} \hat s^j
\eqno(3.3)
$$
Then the cross-section for the production of $N$ gluon
partons plus 2 quark partons is
$$
\sigma ^{(N)}(s)={1\over 2\pi i}\int dj\; \sigma ^{(N)}(j)\; \hat s^j$$$$
\sigma ^{(N)}(j)=
{8\over 9} \left ({3\over\pi ^2}\right )^N\int d^2k_0d^2k_1\dots d^2k_N~~~~~~~~~~~~~~~~~~~~~~~~~~~~~~~~~
~~~~~~~~~~~~~~~~~~~~~~$$$$
~~~~~~~~~~~~~~~~~~~\Psi _j(k_0^2)\Psi _j(k_1^2)\dots\Psi _j(k_N^2)
{\alpha _s\over k_0^2}{\alpha _s\over (k_0-k_1)^2}{\alpha _s\over (k_1-k_2)^2}
\dots{\alpha _s\over(k_{N-1}-k_N)^2}{\alpha _s\over k_N^2}
\eqno(3.4a)
$$
with
$$
\Psi _j(k^2)=\left [j+\half\phi(k^2)\right ]^{-1}
\eqno(3.4b)
$$

Introduce the variables 
$$
K_0=k_0,\;\; K_1=k_0-k_1,\;\;
K_2=k_1-k_2,\;\;\dots ,\;\; K_N=k_{N-1}-k_{N},\;\; K_{N+1}=k_N
\eqno(3.5)
$$
which are just the transverse momenta of the partons -- see the kinematics
shown in figure 3. We impose the conditions that each parton has transverse
momentum at least equal to $\mu$, and that the total transverse energy
of the partons is no more than $\surd s$:
$$
K_r^2 > \mu ^2 ~~~~~~~~~~~~r=0,1,2\dots N+1$$$$
|K_0|+|K_1|+|K_2|+\dots +|K_{N+1}|<\surd s
\eqno(3.6)
$$

Because $\phi$  represents virtual-gluon insertions, the integration (3.2)
should  not have an upper limit. Nor should it have a lower limit: it does
not make sense simply to remove the nonperturbative region from the integration.
We have to decide what to take for the argument of $\alpha _s$; once this is
done, the large-$k^2$ behaviour of $\phi (k^2)$ is determined, independently
of how one handles the nonperturbative region.
We choose to make $\alpha _s$ run with $k^2$ - the scale of $\alpha _s$ can only
be determined by a nonleading calculation, hence we can use the most convenient
choice; then for large $k^2$
we find that $\phi$ becomes constant:
$$
\phi (k^2)\to 2C={72\over  33-2N_f}
\eqno(3.7)
$$
A well-motivated way\footnote{a}{This is very similar
to the treatment found in 
ref. 5, which introduces a gluon mass to regulate the infrared divergences at 
intermediate steps of the calculation. In our case, the introduction of a small
mass has no effect on the real emissions since $ m<<\mu$, but does matter in the
virtual terms, which do depend on the details of the infrared region. 
} to handle the nonperturbative region is that of
Cornwall\defref\cornwall{
J M Cornwall, Physical Review D26 (1982) 1453
}
who deduced by solving Schwinger-Dyson equations that the gluon
propagator $D(q^2)$ and the running coupling should be well approximated by
$$
D^{-1}(q^2)=q^2+m^2(q^2)$$$$
\alpha _s(k^2)= {12\pi\over (33-2N_f)
\log \left [{k^2+4m^2(k^2)\over \Lambda ^2}\right ]}
\eqno(3.8a)
$$
where the running gluon mass is given by
$$
m^2(q^2)=m^2\left [{\log {q^2+4m^2\over\Lambda ^2}
\over\log{4m^2\over\Lambda ^2}}\right ]^{-12/11}
\eqno(3.8b)
$$
The fixed mass $m$ is determined\defref\cond{
A Donnachie and P V Landshoff, Nuclear Physics B311 (1988/9) 509
}\defref\halzen{ 
M B Gay Ducati, F Halzen and A A  Natale,    
Physical Review D48 (1993) 2324
} 
from the condition that the simple exchange of a pair of gluons between
quarks is the zeroth-order approximation to
soft pomeron exchange at $t=0$\footnote{b}{ 
Note that the value of this mass is an
intrinsic QCD parameter, which comes from the structure of the vacuum, 
and that it is in no way related to $ \mu$.}
This requires that the integral
$$
\beta _0^2={4\over 9}\int d^2k\;[\alpha _s(k^2) D(k^2)]^2
$$
be about 4 GeV$^{-2}$. With the choice $\Lambda=200$ MeV this gives
$m=340$ MeV. 

A consequence of $m$ being quite small is that $\phi (k^2)$,
which is now given by
$$
\phi (k^2)={3\over\pi ^2}\int d^2q\;\alpha _s(k^2) k^2 D(q^2)D((k+q)^2)
\eqno(3.2b)
$$
rises rapidly from its value 0 at $k^2=0$ and is already close to its
asymptotic value by $k^2=10$ GeV$^2$. We show this
in figure 4,   where we have taken 4 flavours, so the asymptotic value
(3.8b) is 2.88. 
\topinsert
\centerline{{\epsfxsize=130truemm\epsfbox{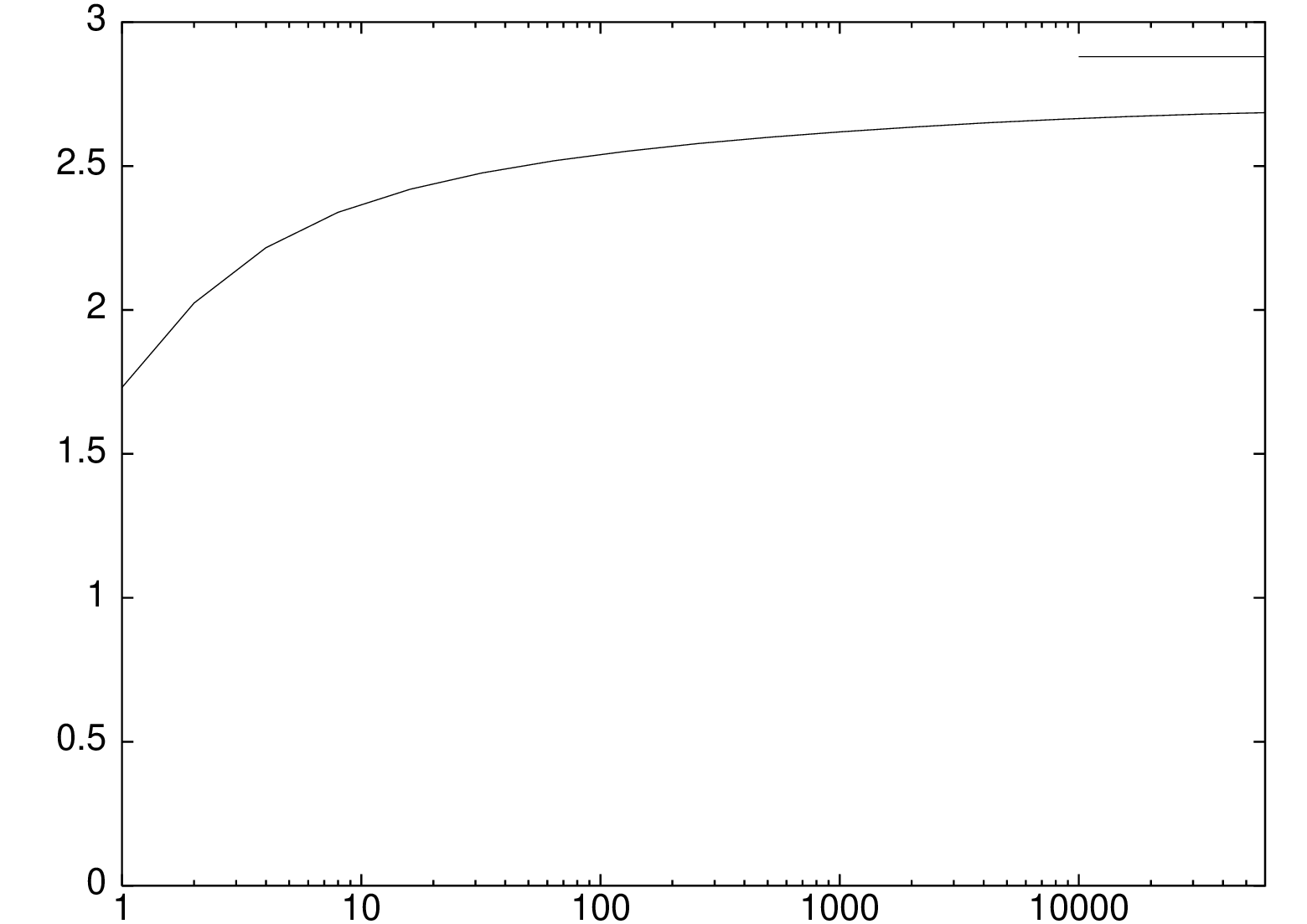}}}\hfill\break
\vskip -10mm
\+&&&&&$k^2$ (GeV$^2$)\cr
\vskip 9pt
\centerline{Figure 4: The function $\phi$; the horizontal line denotes the
asymptotic value}
\endinsert
\def\C{C_{{\hbox{\sevenrm eff}}\, }}

The infrared cut-offs (3.6) on the variables $K$ will tend to suppress
contributions from small values of the variables $k$ also. 
Hence it should be a good numerical
approximation to assign $\phi (k^2)$ a 
constant value $2~\C$
somewhere in the range 1.5 to 2.88. 
We discuss this in section 7;
without such an approximation, further calculation is very
difficult. The larger the value of $\C$, the smaller the output, so if
we are trying to estimate an upper bound to the amplitude we should
take a fairly small value for $\C$. We shall work with $\C$=1.0. We
discuss this choice in section 7. With constant $\C$,
$$
\sigma ^{(N)}(j)={8\pi ^2\over 27} \left({3\over\pi ^2 (j+\C)}\right )^{N+1}
\int d^2K_0d^2K_1\dots d^2K_{N+1}~~~~~~~~~~~~~~~~~~~~~~~~~~~~~~~~~~~$$$$
~~~~~~~~~~~~~~~~~{\alpha _s\over K_0^2}{\alpha _s\over K_1^2}\dots
{\alpha _s\over K_{N+1}^2}
\delta ^2(K_0+K_1+\dots +K_{N+1})
\theta (\surd s-|K_0|-|K_1|-\dots -|K_{N+1}|)
\eqno(3.9)
$$
Introduce the representations
$$
\delta ^2(\kappa)={1\over 4\pi ^2}\int d^2b\; e^{ib.\kappa}$$$$
\theta (\surd s-E)\;\theta (E)={i\over 2\pi}
\int dc\; e^{icE}\;{e^{-ic\surd s}-1\over c}
\eqno(3.10)
$$
We can then sum over $N$, with the result that the part of the cross-section
where the final state contains only partons with transverse momentum
greater than $\mu$ is
$$
\sigma (s|K_T>\mu)
={i\pi\over 81}\hat s^{\,-\C}\int dc\; d^2b\; {e^{-ic\surd s}-1\over c}
[I(b,c)]^2 \hat s^{I(b,c)}
\eqno(3.11a)
$$
with
$$
I(b,c)={3\over\pi ^2}\int d^2K\;{\alpha _s(K)\over K^2}e^{i(b.K+c|K|)}
$$$$=C\int _{\mu}^EdK\;{1\over K\log(K/\Lambda)}J_0(bK)
e^{icK}
\eqno(3.11b)
$$
with $C$ as in (3.7).
Here, somewhat arbitrarily, we have chosen to make $\alpha _s$ run with
$K$. We have also introduced the cut-offs of (3.6). The
upper limit $E$ on the $K$ integration in (3.11b) can be any value not less
than $\surd s$;
the $\theta$-function in
(3.9) ensures that values of $K$ greater than $\surd s$ will not contribute.

Numerical integration of (3.11) shows that, not surprisingly, the result is
very sensitive to the value chosen for $\mu$. This is partly because
the running coupling is largest at the lower end of the $K$ integrations.
We use the lowest-order $\alpha_s$, so that
$\alpha_s(\mu )= 0.33$ at $\mu =2$ and 
0.47 at $\mu =1$.
We believe that the latter value, at least, is too large for a perturbative
calculation of the cross-section to be valid: for
reasonable safety, we should choose $\mu$ to be at least 2. 
(We recall that the usual evaluation of the simple
answer (2.10) for the Lipatov power chooses $\alpha _s$ to be just less than
0.2, making the power 0.5.)

Although the BFKL pomeron is not supposed to be the dominant term in 
total 
cross sections, 
we can certainly calculate the perturbative contribution to $pp$ and
$\bar pp$ once we impose that the intermediate state lies in the 
perturbative region.  Surprisingly, we find that we have
to go to rather large values of $\mu$ to ensure compatibility with
the data.

The curve in figure 5
shows the output for $\sigma (s|K_T>\mu)$ with $\mu =2$~GeV.
The output shown
is for quark-quark scattering; according to the additive-quark rule we should
multiply it by 9 to obtain the contribution to the $pp$ or $\bar pp$
total cross-section. However, the variable $\surd s$ is the centre-of-mass
energy of the quark pair, which is approximately 1/3 that of the
$pp$ or $\bar pp$ system.
(We are assuming that the quark additivity
that is valid\ref{\total} for soft pomeron exchange is also applicable
to hard-pomeron exchange. This means that we are neglecting a possible 
shadowing suppression associated with the gluons coupling to different
quarks in a hadron.)
\topinsert
\centerline{{\epsfxsize=120truemm\epsfbox{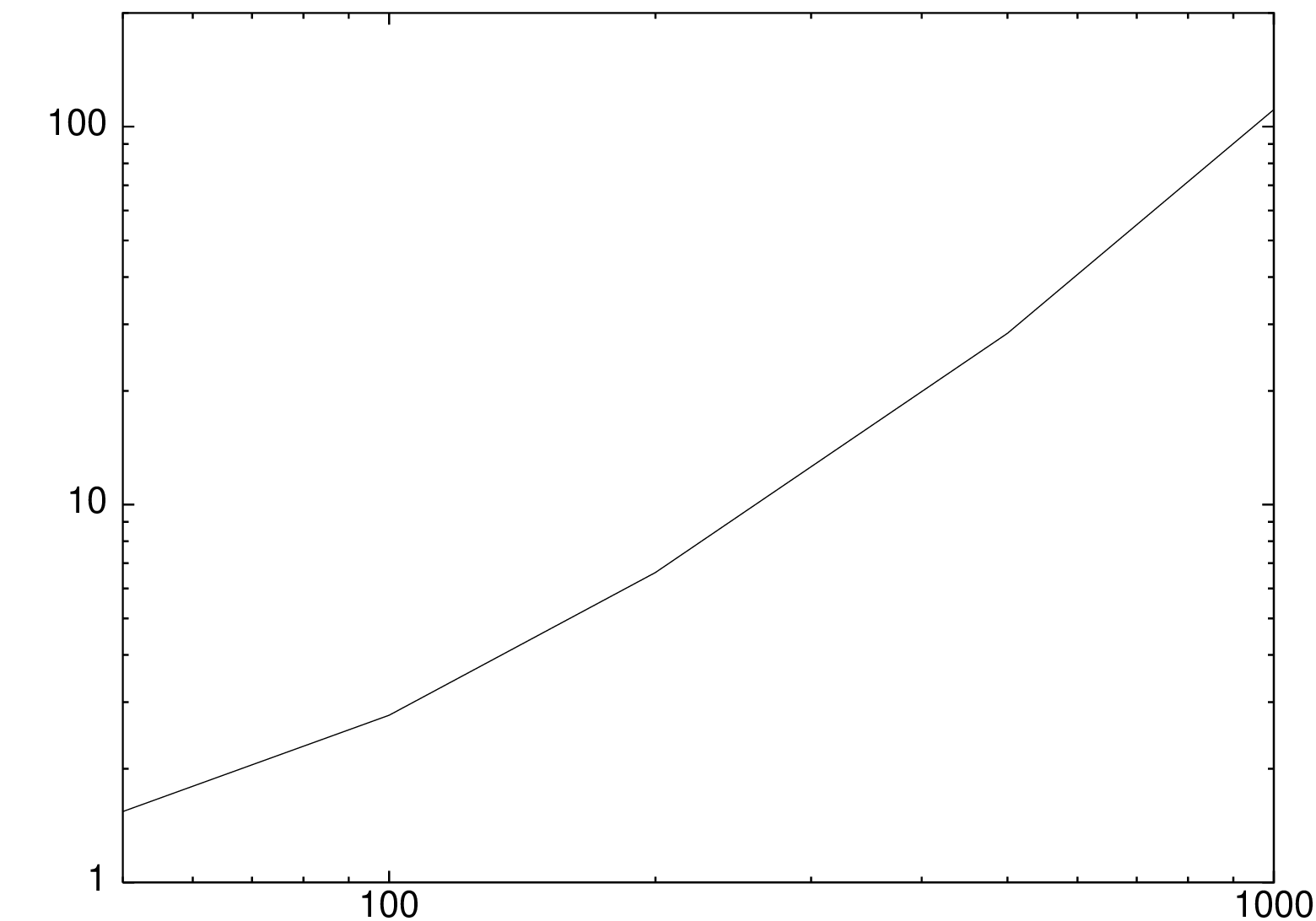}}}\hfill\break
\vskip -10mm
\+&&&&&$\surd s$ (GeV)\cr
\centerline{Figure 5: $\sigma _{qq}(K_T>\mu)$ in microbarns for $\mu =2$ GeV} 
\endinsert

We show in the next section that $\sigma (s|K_T>\mu)$ must be multiplied
by a factor which may approach an order of magnitude, to account for
nonperturbative effects. 
We may conclude from $pp$ or $\bar pp$ total cross-section data that
the value  2 GeV we have used for $\mu$ is the minimum safe value 
for the perturbative calculation. This is because
there is little or no room
in the existing $pp$ or $\bar pp$ total cross-section data for anything
that rises with $s$ more rapidly than soft pomeron exchange\defref\margolis{
J R Cudell and B Margolis, Physics Letters B297 (1992) 398
}.  We recall\ref{\total} that soft-pomeron 
exchange describes the rising component
of the cross sections extremely well over a huge range of $s$, from
$\surd s=5$ GeV or less, to 1800~GeV. There perhaps is some room 
in the data for a hard-pomeron component in addition,
depending on which of the two conflicting Tevatron
experiments\defref\tev{
E710 collaboration: N Amos et al, Phys Lett B243 (1990) 158\h
CDF collaboration: F Abe et al, Physical Review D50 (1994) 5550
} one believes. If one accepts the CDF result, there could be a
hard-pomeron contribution that has reached as much as 10 mb at Tevatron
energy. Because it would fall rapidly with decreasing $\surd s$,
this would not cause a problem with the fit to the data at ISR
energies and below. 10~mb is approximately 
the value we deduce from figure 5 when we allow for nonperturbative
corrections. If we changed from $\mu=2$ GeV to 1 GeV, we would obtain at
$qq$ energy
$\surd s=600$~GeV an increase of 3 orders of magnitude, which is certainly
excluded. If the hard-pomeron-exchange contribution at the Tevatron is
actually somewhat less than 10 mb, then the message is that the ``safe''
value for $\mu$ is higher than 2 GeV.

Before we discuss
the nonperturbative corrections to the perturbative
calculation in the next section,
we point out the importance of the energy-conservation constraint.
Without this constraint, (3.11) becomes
$$
\sigma (s|K_T>\mu)
={2\pi ^2\over 81}\hat s^{\,-\C}\int d^2b\
[I(b,0)]^2 \hat s^{I(b,0)}
\eqno(3.12a)
$$
with
$$
I(b,0)=C\int _\mu ^\infty dK\;J_0(bK){1\over K\log (K/\Lambda)}
=C\int _{b\mu}^\infty dz\;J_0(z){1\over z\log (z/b\Lambda)}
\eqno(3.12b)
$$
Figure 6 shows the ratio of (4.1) to (3.11a) for the case $\mu=2$ GeV.
\topinsert
\centerline{{\epsfxsize=120truemm\epsfbox{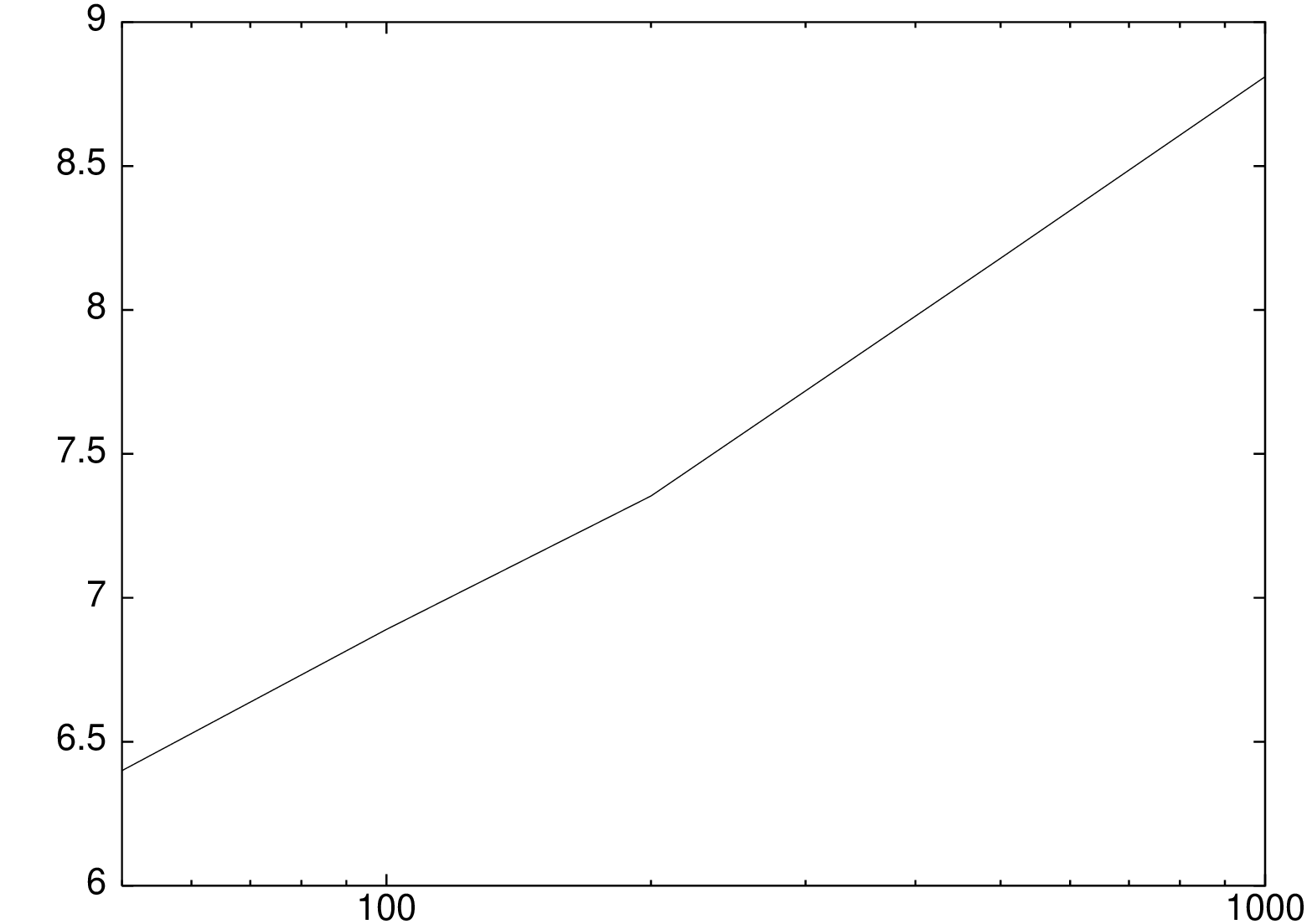}}}\hfill\break
\vskip -10mm
\+&&&&&$\surd s$ (GeV)\cr
\vskip 9pt
\centerline{Figure 6: Effect of energy conservation: ratio of (4.1) to
(3.11a) for the case $\mu=2$ GeV}
\endinsert

\bigskip
\goodbreak
{\bf 4 Complete total cross-section}

In section 3 we derived $\sigma_{qq}{(K_T>\mu)}$, the
contribution to the total quark-quark cross-section from events
where the final state contains only partons with transverse 
momentum greater than $\mu$. We showed that,
while this has the expected fierce variation with energy,
it is numerically 
quite small at reasonable energies.  In this section we explore whether
the fiercely-varying contribution is significantly enhanced by the 
inclusion of final states where there are also partons whose
transverse momentum is less than $\mu$. Our calculation is for a
``quenched'' pomeron: we do not include quark loops, though they may well
be important\defref\quarkloop{
A Donnachie and P V Landshoff, Physics Letters B202 (1988)  131
}.

Define $\sigma_{qq} '(s)$ by
$$
\sigma_{qq}=\sigma_{qq} ' + \sigma_{qq}{(K_T>\mu)}
\eqno(4.1)
$$
The choice of $\mu$ is arbitrary, except we want to make it large enough
for us to be able to calculate $\sigma_{qq}{(K_T>\mu)}$ purely
perturbatively. Because $\sigma_{qq}$ is the total cross-section, it does
not depend on $\mu$, so $\sigma_{qq} '$ must have a variation with $\mu$ that
compensates that of $\sigma_{qq}{(K_T>\mu)}$. However, this
variation is a tiny fraction of the total $\sigma_{qq} '$, because 
$\sigma_{qq}{(K_T>\mu)}\ll\sigma_{qq}$. Thus, while by
definition $\sigma_{qq}{(K_T>\mu)}$ is the result of the
exchange of the perturbative BFKL pomeron $\P _H$, $\sigma_{qq} '$ essentially
corresponds to the observed exchange of the soft pomeron $\P _S$.
In what follows, we also need
$\sigma_{qq} (K_T<\mu)$, the contribution to the cross-section from events where there
are no partons having transverse momentum greater than $\mu$.
Evidently $\sigma_{qq} (K_T<\mu)\leq\sigma_{qq} '$ and 
therefore $\sigma_{qq} (K_T<\mu)\approx\sigma_{qq}$
provides a good estimate of the largest possible value for 
$\sigma_{qq} (K_T<\mu)$. 
Of course, at reasonably high energy it is rather likely that at least
one parton with transverse momentum greater than $\mu$ will be produced,
so that actually $\sigma_{qq} (K_T<\mu)$ will be considerably less 
than $\sigma_{qq}$.

Exactly similar statements apply to the cross-sections for quark-gluon
and gluon-gluon scattering. Also, these cross-sections are related by
factorisation to those for quark-quark scattering. In the case of those 
cross-sections that correspond to the exchange of the
soft pomeron $\P _S$, in particular our
upper estimate for $\sigma (K_T<\mu)$, this factorisation is the result of
the pomeron apparently being a simple pole in the complex angular momentum 
plane; it has recently been well tested\ref{\diff} at HERA. In
the case of the exchange of the BFKL pomeron $\P _H$, the factorisation 
results from
the factorisation of the leading part of the lowest-order contribution:
the coupling to a gluon, averaged over spins and colours, is 9/4 times
that to a quark\defref\combridge{
B L Combridge, Physics Letters 70B (1977) 234
}.

We explained in section 1 how,
in a general event, we may group  the final-state partons according to
their rapidities, and that summing over final states leads to
the series (1.2). 
As we have explained above, by using the full soft pomeron $\P _S$
we over-estimate $\sigma (K_T<\mu)$ and then have upper bounds for
the terms in (1.2). 
The question we explore now is whether, by coupling $\P _H$ to quarks
through $\P _S$ instead of directly, we increase the effective strength
of its coupling. First, we investigate whether the term
$\P _S\otimes\P _H\otimes\P _S$ is significantly larger than $\P _H$.
We show this term in figure 7.
In a previous paper\defref\minijets{
A Donnachie and P V Landshoff, Physics Letters B332 (1994) 433
}, 
we have calculated the term in the approximation where $\P _H$ is
replaced with its lowest-order contribution. It is a simple matter to
use instead the whole of $\P _H$:
$$
\P _S\otimes\P _H\otimes\P _S=32\pi ^2\beta _0^2g^2 
\int _{4\mu ^2}^s {ds_{12}\over s_{12}}
\left ({s\over s_{12}}\right )^{\epsilon _0}
\sigma_{gg}(s_{12}|K_T>\mu) \log\left ({s\over s_{12}}\right ) 
\eqno(4.2)
$$
This is for quark-quark scattering. Here, $(1+\epsilon _0)$ with
$\epsilon _0\approx 0.08$ is the intercept of the soft pomeron,
$\beta _0\approx 2$ GeV$^{-1}$ is the strength of its coupling to a
quark, and $g\approx 15$ MeV its coupling to a gluon. (We defined\ref{\minijets}
$\beta _0$ and $g$ in somewhat different ways, such that they even 
have different dimensions, so the values we have
given for them do not directly reflect the relative strengths of the two
couplings). We fit the upper energy part of the
curve in figure 5 with an effective power,
$$
\sigma_{qq}{(s|K_T>\mu)}\sim 2\beta _H^2 (s/\mu ^2)^{\lambda}
\eqno(4.3a)
$$
This gives
$$
\lambda\approx 0.86$$$$ 
\beta _H^2\approx 3.3\times 10^{-6} {\hbox{GeV}}^{-2}
\eqno(4.3b)
$$
Then $\sigma_{gg}(s_{12}|K_T>\mu)$ is 81/16 times this, 
We find that (4.2) is approximately
$$
{324\pi ^2\beta _0^2g^2\beta _H^2\over
(\lambda -\epsilon _0) ^2}(s/\mu ^2)^{\lambda}
\approx {1.5\over \lambda ^2}\sigma_{qq}{(s|K_T>\mu)}
\eqno(4.4)
$$
This is a few times $\sigma_{qq}{(s|K_T>\mu)}$, but remember that it is
an upper bound. 
\topinsert
\centerline{{\epsfxsize=30truemm\epsfbox{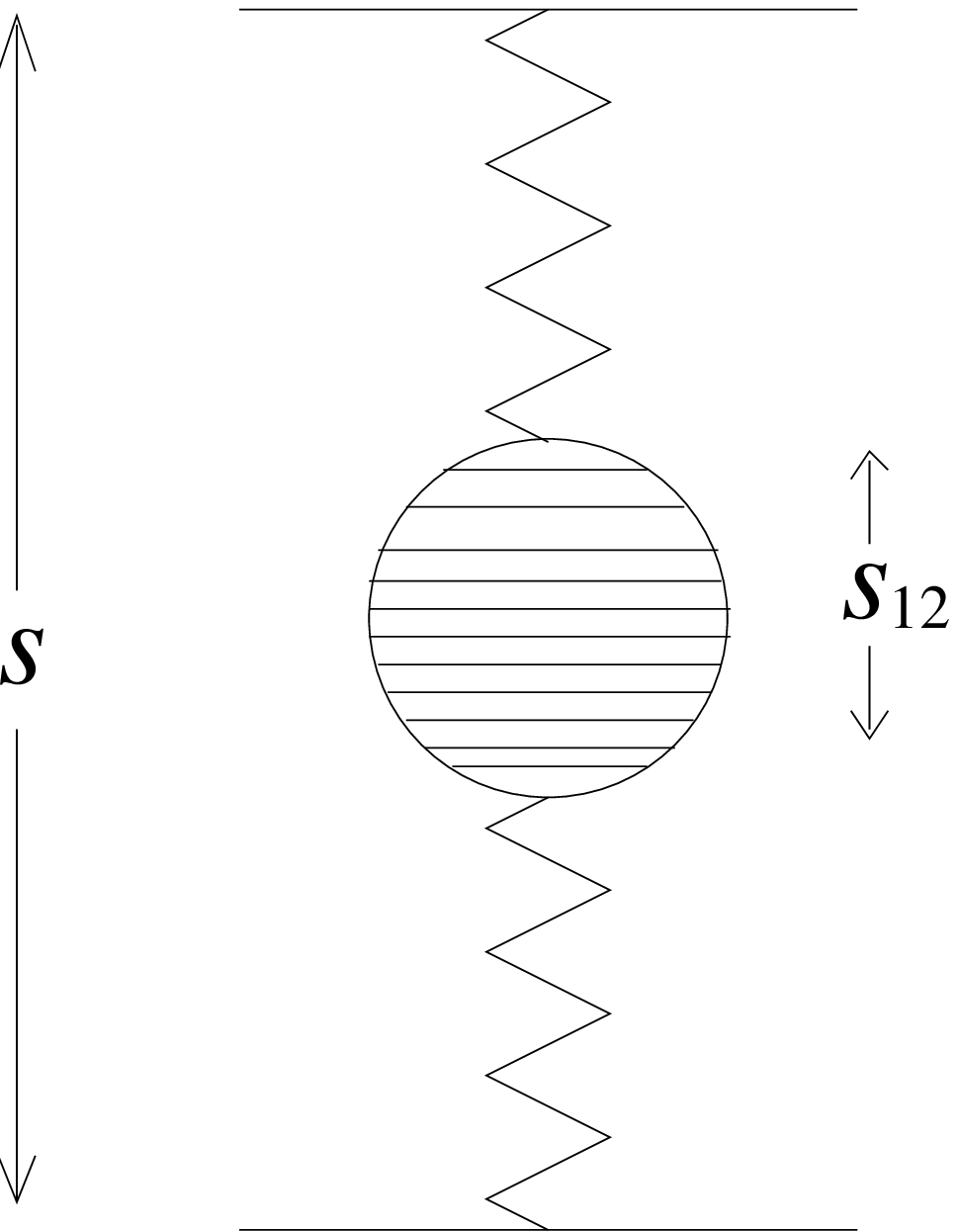}}}\hfill\break
\centerline{Figure 7: Hard exchange sandwiched between two soft exchanges}
\endinsert

Consider now the term $\P _S\otimes\P _H\otimes\P _S\otimes\P _H\otimes\P _S$,
where the final state contains two groups of high-transverse-momentum
partons. This term is\ref{\minijets}
$$
512\pi ^4\beta _0^2g^4\int _{4\mu ^2}^s {ds_{12}ds_{34}\over s_{12}s_{34}}
\left ({s\over \alpha ' s_{12}s_{34}}\right )^{\epsilon _0}\;
\sigma_{gg}(s_{12}|K_T>\mu)\sigma_{gg}(s_{34}|K_T>\mu)\;
\log ^2L \;\theta (L-1)
$$$$
L={\mu _0^2s\over s_{12}s_{34}}
\eqno(4.5)
$$
Here, the upper limit on the integrations is the surface $L<1$ and
$\mu _0$ is a nonperturbative scale, expected\ref{\minijets}
to be about 1 GeV, associated with the coupling of the soft pomeron to gluons.
Using again the effective-power description (4.4) for $\sigma_{gg}(K_T>|\mu)$,
we find that (4.5) gives (4.4) times
$$
{324\pi ^2g^2 \beta _H^2\over (\lambda -\epsilon _0)}
(\mu ^2/\mu _0^2)^{\lambda}\log{\mu _0^2s\over 16\mu ^4} 
\eqno(4.6a)
$$
which is approximately (4.4) times
$$
{2\times 10^{-6}\over \lambda}
(\mu ^2/\mu _0^2)^{\lambda} \log{\mu _0^2s\over 16\mu ^4}
\eqno(4.6b)
$$
This is
much less than (4.4) until the energy is very
high indeed.  

We must consider also the terms $\P _S\otimes\P _H$ and $\P _H\otimes\P _S$.
They are equal, and their sum works out to be
$$
36\pi g\beta _0\sigma_{qq}{(s|K_T>\mu)}
\approx 3 \sigma_{qq}{(s|K_T>\mu)}
\eqno(4.7)
$$
Again this is an upper bound.

Our conclusion is that the inclusion of the nonperturbative contributions
multiplies the rapidly-rising component of the cross section
by a number that is at is at most an order
of magnitude when the lower limit $\mu$ of the perturbative calculation
is chosen to be 2 GeV. As we explained in the last section, this leads
us to conclude that it is unsafe to choose $\mu$ to be significantly lower
than this, because it would conflict with data for the $\bar pp$
total cross section. 
\bigskip
{\bf 5 Semihard processes}

As a first look at a semihard process, we consider $\gamma ^*q\to\rho q$.
At high $Q^2$, the dominant polarisations for the $\gamma ^*$ and $\rho$
are longitudinal\defref\rrho{
A Donnachie and P V Landshoff, Physics Letters B185  (1987) 403
and B348 (1995) 213}\defref\jean{
J R Cudell, Nuclear Physics B336 (1990) 1\h
J M Laget and R Mendez-Galain, Nuclear Physics A581 (1995) 397
}\ref{\cond}.
In lowest order, the amplitude 
is given
by the graph of figure 8. We consider forward scattering in the zero-mass
limit. The amplitude is
$$
a_0(Q^2)= i\int {d^2k_T\over k_T^4}
u _0(Q^2,k_T^2)$$$$
u _0(Q^2,k_T^2)={16 e  f_{\rho}
\over 3\surd 3 Q}{\alpha _s^2k_T^2\over (k_T^2+\quarter Q^2)} 
\eqno(5.1)
$$
We have used a nonrelativistic wave function for the $\rho$.
We have used perturbative gluon propagators, in place of the nonperturbative
ones of reference {\cond}, because here we wish to  consider the
exchange  of the perturbative pomeron rather than the nonperturbative.
The similarity with (2.3) is evident, so that we may immediately write
down the result of making all the higher-order insertions as in sections
2 and 3. Instead of (3.11a), 
$$
a(s,Q^2|K_T>\mu)=
{2\pi e f_{\rho}
\over 27\surd 3 Q}\;
x^{\C}\int dc\; d^2b\; {e^{-ic\surd s}-1\over c}
I(b,c)J(b,c,Q^2)  x^{-I(b,c)}
\eqno(5.2)
$$
with
$$
J(b,c,Q^2)={3\over\pi ^2}\int d^2K\;{\alpha _s(K)\over K^2+
\quarter Q^2}
e^{i(b.K+c|K|)}
\eqno(5.3)
$$
We assume that it is now appropriate to replace  $\hat s$ with 
$x^{-1}=s/Q^2$.
The forward-scattering differential cross-section is obtained by squaring
the amplitude (5.2), and dividing by $16\pi^2$.
We should have to multiply by 3 if we were to change from
a quark target to a proton. 
\topinsert
\centerline{{\epsfxsize=120truemm\epsfbox{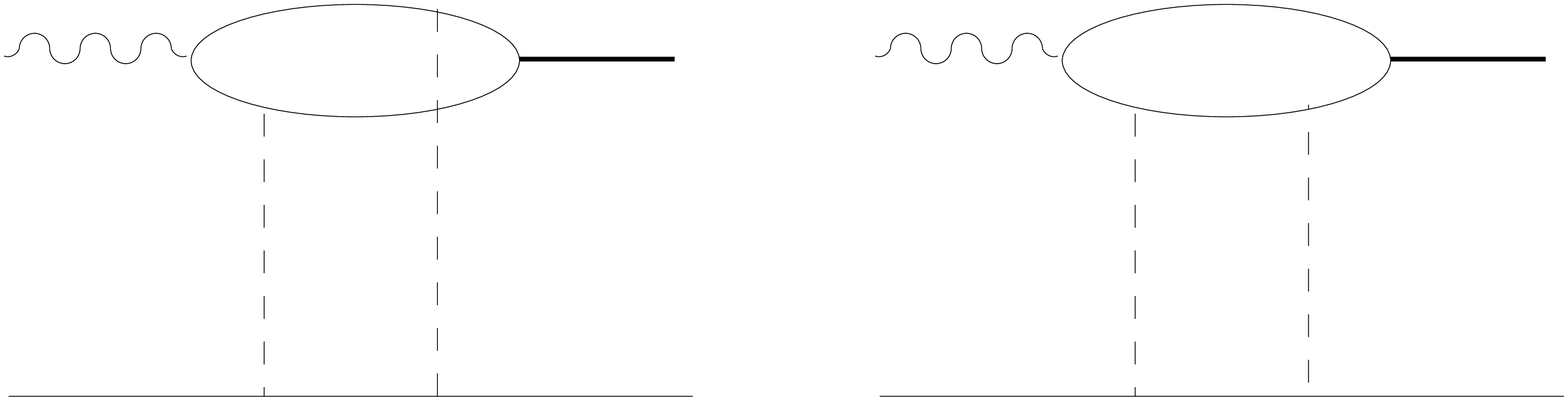}}}\hfill\break
\centerline{Figure 8: Lowest-order graphs for $\gamma ^*q\to\rho q$}
\endinsert

It seems that, for large enough $Q^2$, the amplitude (5.2) at fixed $x$ varies
as 1/$Q^3$. However,
it turns out that the integral varies quite slowly with
$Q^2$ until $Q^2$ becomes extremely large, and so until then the fall-off
with increasing $Q^2$ of the differential cross-section is much
slower than $1/Q^6$. 
This conclusion contrasts with that of Brodsky and
collaborators\defref\brodsky{
S J Brodsky et al, Physical Review D50 (1994) 3134
}, 
who guess that the asymptotic $Q^2$-dependence of perturbative
exchange is achieved quite early.  

At NMC energies,  it is soft pomeron exchange
rather than perturbative exchange that describes the data\ref{\rrho}%
\defref\nmc{
NMC collaboration: P Amaudraz et al, Z Physik C54 (1992) 239;\h
M. Arneodo et al, Nuclear Physics B429 (1994) 503 
}.
This has been tested out to the $Q^2 \approx 20$ GeV$^2$, by which time
the soft exchange has already achieved the $1/Q^6$ fall-off. 
But at HERA energies for the same 
$Q^2$ values there seems
to be a more rapid rise\ref{\rise}
with energy than is expected from soft pomeron exchange.
One might seek to explain this by supposing that the BFKL contribution
is fairly small at NMC energies but, with its more rapid rise than
the soft pomeron term, it has become important at HERA energies. 
Our calculations suggest that
this is unlikely. 

The soft-pomeron-exchange amplitude is\ref{\rrho}
$$
a_{{\rm soft}}(s,Q^2)= \left ({s\over s_0}\right )^{\epsilon _0}
\int {d^2k_T\over (k_T^2+\quarter Q^2)}k_T^2 D^2(-k_T^2)
u _0(Q^2,k_T^2)
\eqno(5.4a)
$$
where $D$ is the nonperturbative gluon propagator\ref{\otto}
and $\epsilon _0\approx 0.08$
is the intercept of the soft pomeron trajectory, with $s_0\approx 4$ GeV$^2$.
For large $Q^2$ the soft  amplitude may be written as\ref{\rrho}
$$
a_{{\rm soft}}(s,Q^2)={8\surd 3 i e f_{\rho}\over Q^3}\beta _0^2\mu _0^2
\left ({s\over s_0}\right )^{\epsilon _0}
\eqno(5.4b)
$$
where $\beta _0$ is the coupling of the soft pomeron to a light quark
and $\mu _0$ is a mass scale which experiment finds\ref{\rrho} to be about
$2/\beta _0$.
\topinsert
\centerline{{\epsfxsize=140truemm\epsfbox{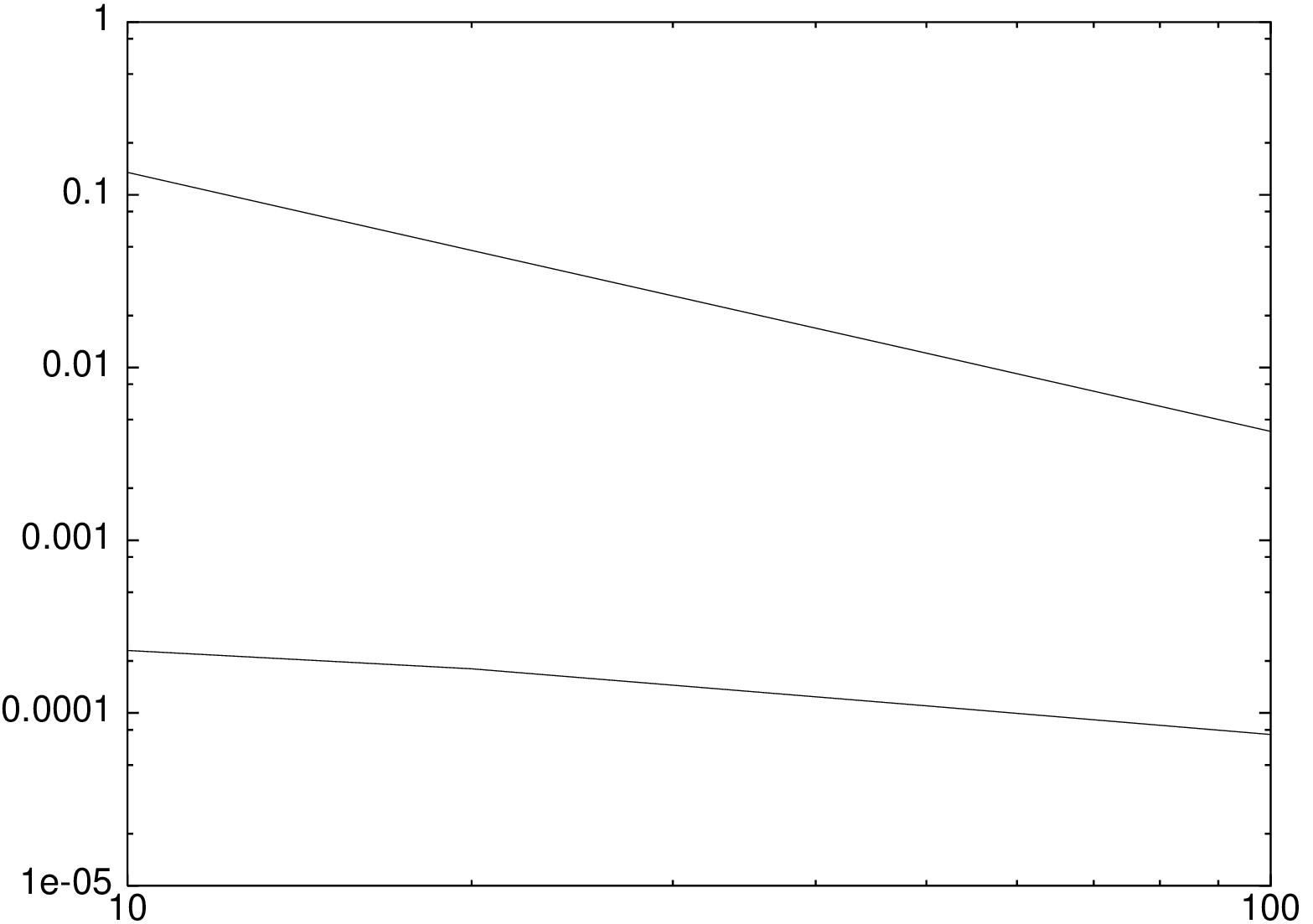}}}
\+&&&&&$Q^2$ (GeV$^2$)\cr
\+&&&(a)\cr
\vskip 10pt
\centerline{{\epsfxsize=140truemm\epsfbox{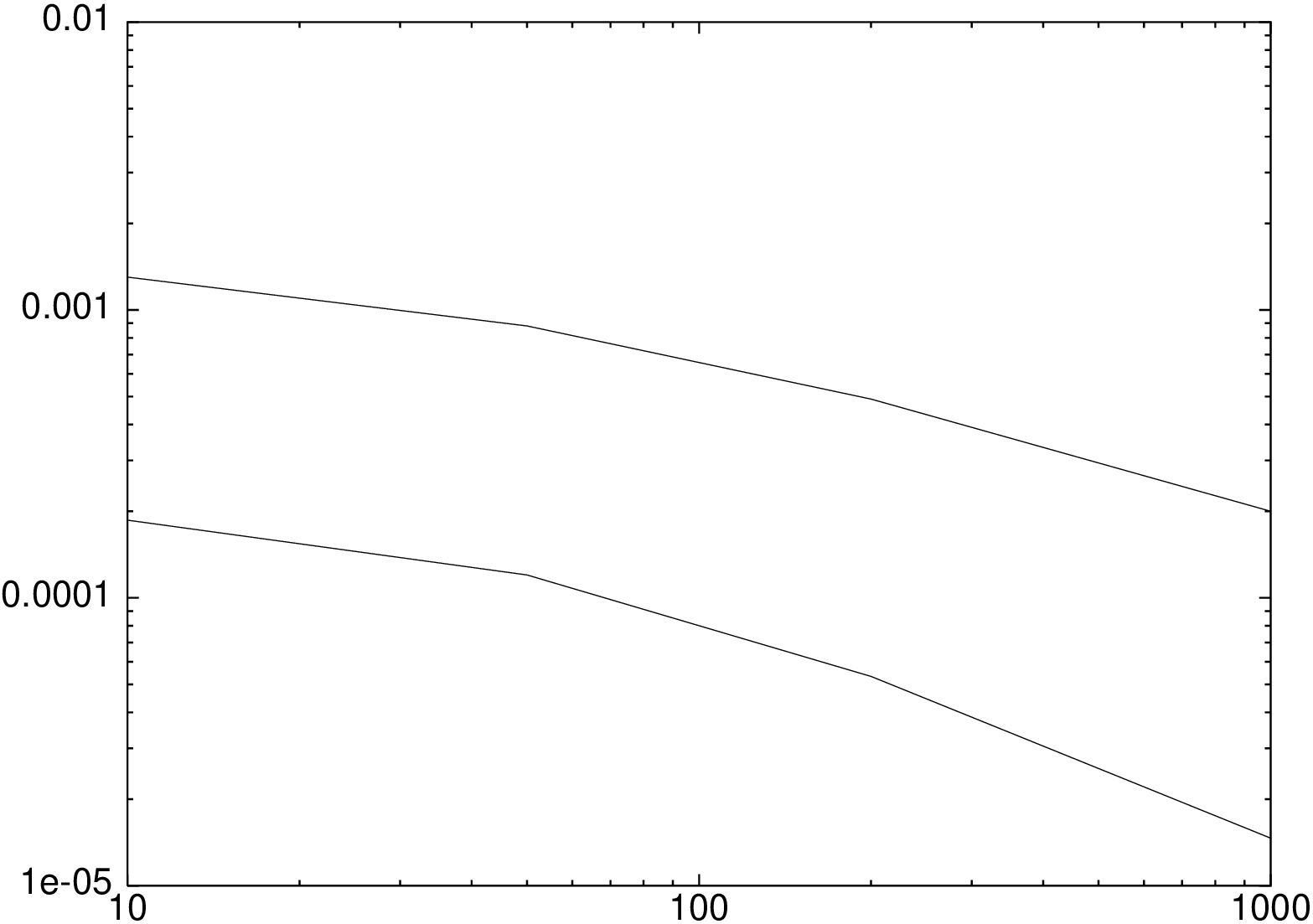}}}\hfill\break
\vskip -10mm
\+&&&&&$Q^2$ (GeV$^2$)\cr
\+&&&(b)\cr
{Figure 9: the amplitude for $\gamma ^*q \to\rho q$ in GeV units 
(a)   at $\surd s=50$ GeV, hard exchange 
(lower curve) and soft exchange (upper curve)
(b)  hard exchange at $x=0.01$ (lower curve) 
and $x=10^{-4}$ (upper curve)}
\endinsert
\topinsert
\centerline{{\epsfxsize=140truemm\epsfbox{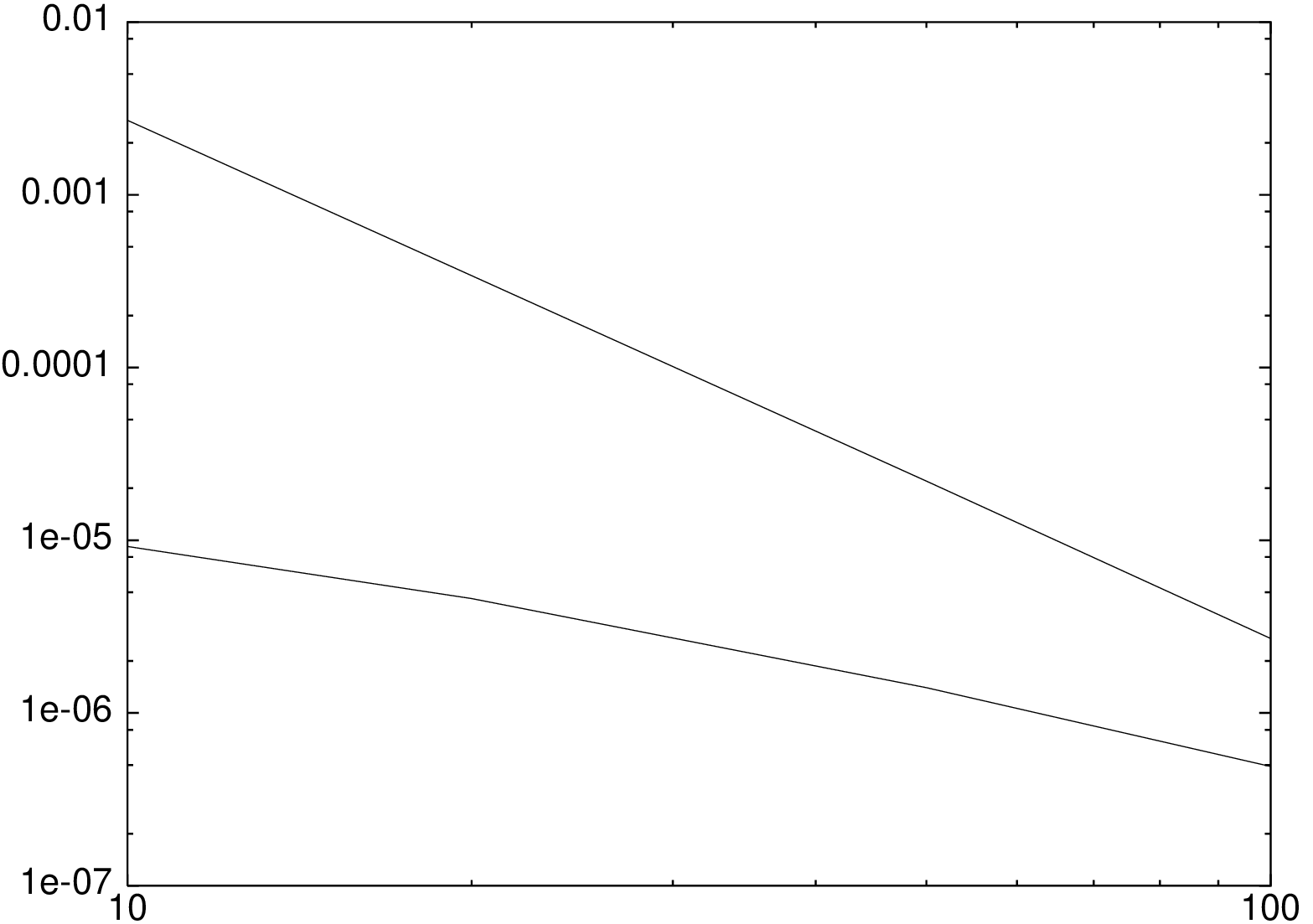}}}
\+&&&&&$Q^2$ (GeV$^2$)\cr
\+&&&(a)\cr
\vskip 10pt
\centerline{{\epsfxsize=140truemm\epsfbox{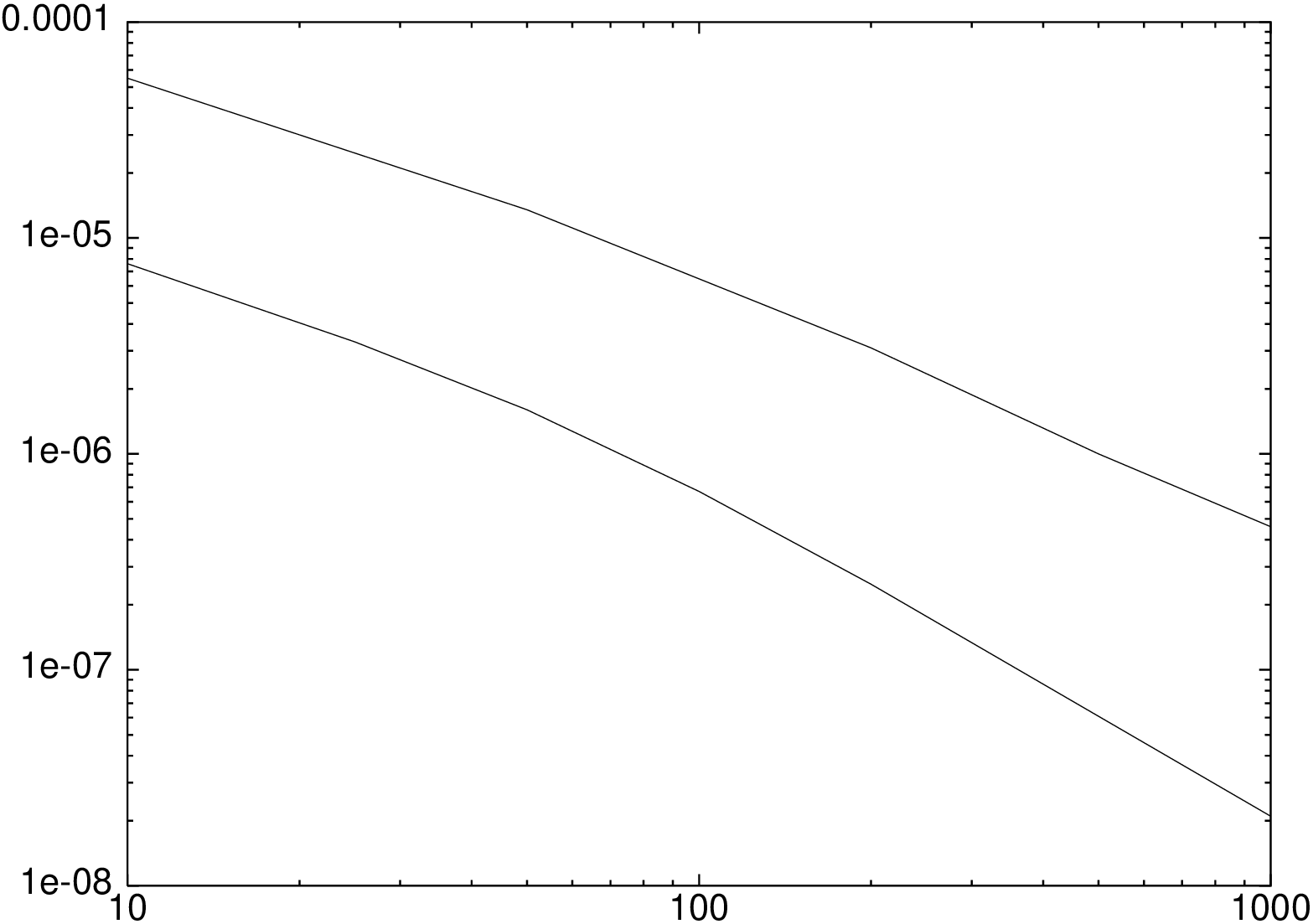}}}\hfill\break
\vskip -10mm
\+&&&&&$Q^2$ (GeV$^2$)\cr
\+&&&(b)\cr
{Figure 10: the amplitude for $\gamma ^*\gamma ^*\to\rho\rho$ in GeV units 
(a)   at $\surd s=50$ GeV, hard exchange 
(lower curve) and soft exchange (upper curve)
(b)  hard exchange at $x=0.01$ (lower curve) 
and $x=10^{-4}$ (upper curve)}
\endinsert

In figure 9a we plot the hard and soft
amplitudes at $\gamma ^*q$ energy $\surd s=50$ GeV. We have used $\mu =2$~GeV.
Since the soft amplitude is of the same order
of magnitude as the data for $Q^2<20$ GeV$^2$, the hard amplitude is obviously
unimportant. Even if we allow for the nonperturbative corrections of
the type we have discussed in section 4, at $Q^2=100$ GeV$^2$ it is
at least an order of magnitude smaller than the soft amplitude.
Figure 9b shows the $Q^2$ variation of the hard amplitude at fixed $x$. 
At $Q^2=1000$ GeV$^2$ the $x$-dependence, averaged between $1/x=100$ and
$1/x=10000$,  is $(1/x)^{0.58}$, but at
smaller $Q^2$ the variation is slower --- $(1/x)^{0.44}$ at $Q^2=10$ GeV$^2$.
Notice that at fixed $x$ the fall-off of the amplitude with $Q^2$ is
very much slower than $1/Q^3$.

In this semihard process the dependence on $\mu$ is found to be much less 
than for the purely soft process. At $\surd s=100$ GeV, changing from
$\mu =2$ GeV to 1 GeV increases the $qq\to qq$
amplitude by two orders of magnitude,
but for the semihard $\gamma ^*q\to\rho q$ at the same $\surd s$ this
factor has reduced to 5 by $Q^2=1000$ GeV$^2$.
\bigskip
{\bf 6 Hard process}

For an initial look at a purely hard process we choose
$\gamma ^*\gamma ^*\to\rho\rho$. We take the two photons to have the
same virtuality $Q^2$ and again consider forward scattering and
longitudinal polarisations. The lowest-order amplitude is 
$$
a_0(Q^2,Q^2)=i\int {d^2k_T\over k_T^4}
v _0(Q^2,Q^2,k_T^2)
\eqno(6.1)
$$
There is $k_T$-factorisation\defref\ginzburg{
I F Ginzburg, S L Panfil and V G Serbo, Nuclear Physics B296 (1988) 569
}, so that
$$
v _0(Q^2,Q^2,k_T^2)={u^2_0(Q^2,k_T^2)\over t_0(k_T^2)}
\eqno(6.2)
$$
and (5.2) becomes
$$
a(s,Q^2,Q^2|K_T>\mu)=
{4\pi e^2f^2_{\rho}\over 27Q^2}
x^{\C}\int dc\; d^2b\; {e^{-ic\surd s}-1\over c}
J^2(b,c,Q^2)  x^{-I(b,c)}
\eqno(6.3)
$$
Superficially, at fixed $x$ the $Q^2$ dependence at large $Q^2$ of this
amplitude is $1/Q^6$, but $Q^2$ needs to be extremely large to get
anywhere near this. It is uncertain how soft-pomeron exchange contributes
to this process, but if we assume that it factorises we obtain the amplitude
$$
a_{{\rm soft}}(s,Q^2,Q^2)={48 i e^2 f^2_{\rho}\over Q^6}\beta _0^4\mu _0^4
\left ({s\over s_0}\right )^{\epsilon _0}
\eqno(6.4)
$$

In figure 10a we plot the hard and soft amplitudes against $Q^2$, 
for $\surd s=50$ GeV. 
If we include the nonperturbative corrections of the type we
discussed in section 4, it may be that the hard amplitude becomes comparable
with the soft by $Q^2=100$ GeV$^2$, but by then the cross-section is
very small.
In figure 10b we plot the hard amplitude at $x=10^{-2}$ and $10^{-4}$.

We said at the end of the last section that for the semihard
process $\gamma ^*q\to\rho q$ at  $\surd s =100$ GeV and $Q^2=1000$ GeV$^2$
the effect of decreasing $\mu$ from 2 to 1 GeV is to increase the
amplitude by a factor of 5, which is very much less than for the purely
soft process $qq\to qq$ at the same energy. For the hard process at the same
$\surd s$ and $Q^2$ the factor is further 
reduced, to about 3. Also, while for the soft process at this energy
relaxing the energy-conservation constraint, as in (3.12), increases the 
amplitude by a a factor close to 7, for the hard process this factor is
only about 5. Further, while for $Q^2=10$ GeV$^2$ the $x$-dependence
averaged between $1/x=100$ and $1/x=10000$ is almost the same as for the
semihard process, at $Q^2=1000$ GeV$^2$ it is rather fiercer:
$(1/x)^{0.69}$ instead of $(1/x)^{0.58}$.
These
conclusions are in line with general expectations about the ``diffusion''
of transverse momentum\ref{\bartels} at high $Q^2$, though the effect
is perhaps not as dramatic as might have been hoped.
\bigskip
{\bf 7 Discussion}

We have argued in this paper that the BFKL pomeron is not detectable,
at least at $t=0$.
This means that some other explanation must be found for the rapid rise
in the HERA data for $\rho$ electroproduction and presumably
also for $F_2$ at small $x$, for example\defref\grv{
M Gl\"uck, E Reya and A Vogt, Z Physik C67 (1995) 433\h
R D Ball and S Forte, Physics Letters B351 (1995) 313
}
the onset of
perturbative Altarelli-Parisi
evolution at a smaller value of $Q^2$ than many people expected.

We have considered only the BFKL pomeron at $t=0$. It remains possible that
at large $t$ the situation will be different since then, even if the
BFKL contribution is small, there is much less competition from 
nonperturbative mechanisms.
We intend examining this in a future paper.

As the discussion of the BFKL pomeron inevitably requires consideration of 
nonperturbative contributions, it must involve considerable uncertainty.
Our approach, when faced with decisions where there is theoretical
uncertainty,  has been to maximise the cross section, within the constraints 
coming from HERA and the Tevatron.
One issue
is the value we have chosen for $\C$.  We have chosen it to be 1.0, and found
in particular that
the perturbative contribution to  $\gamma ^*q\to \rho q$ is totally 
negligible, even if we allow for the type of corrections discussed in section 4.
To make it comparable with the soft contribution (and therefore with the data)
we see from  figure 9a that 
we should need to reduce $\C$ to 0. This is not reasonable, because it 
would correspond to an absence of virtual corrections. 
More importantly, its $x$ dependence would be totally wrong. 
As we show in figure 11,
the choice $\C =1$  already gives an effective power $(1/x)^{0.3}$; 
changing $\C$ to 0 would make this $(1/x)^{1.3}$.
\topinsert
\centerline{{\epsfxsize=120truemm\epsfbox{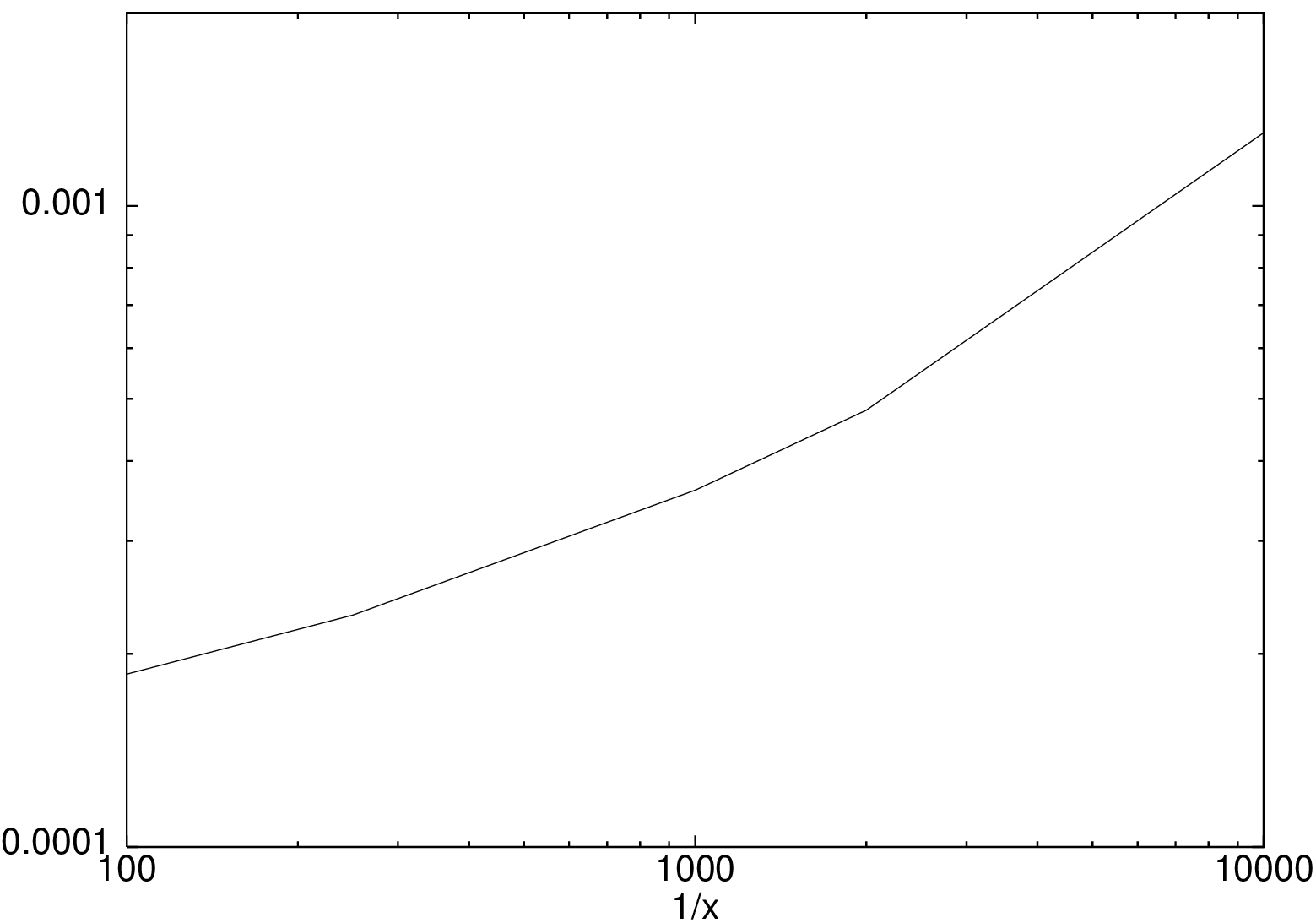}}}\hfill\break
\centerline{Figure 11: the hard amplitude for $\gamma ^*q\to\rho q$
at $Q^2=10$ GeV$^2$}
\endinsert

An alternative way to increase the size of the perturbative contribution is to
reduce the value of $\mu$ (though again this would give the wrong $x$ 
dependence). Also, one might guess
that perhaps $\C$ should vary with $\surd s$, being smaller than our chosen
value 1.0 at the left of figure 5, and larger at the right. 
However,
we have explained that the value of the cross-section at
$qq$ energy $\surd s\approx 600$~GeV 
is constrained by the Tevatron data to be no larger than is shown in
the figure. To bring the hard-exchange curve in figure 9a near to the
level of the soft-exchange curve, we need an increase of at least 2
orders of magnitude at the left in figure 5. If $\C$ is made energy-dependent
in such a way to achieve this, while keeping the curve at 
$\surd s\approx 600$~GeV 
at the same level, the whole curve would become so flat that it
would no longer be distinguishable from soft exchange.

In applying the
analysis to hard and semihard processes, we have used experimental information
from soft processes: that the $pp$ and $\bar pp$ total cross sections have 
at most only a very small very-rapidly-rising component. Although nobody
expects the perturbative contribution to such soft processes to form a
significant fraction of the total, it should nevertheless be present
and it is important to use experimental information about how large it can
be to pin down some of the uncertainties about the corresponding
contributions to semi-hard and hard processes. We have said
that the $pp$ and $\bar pp$ total-cross-section data show that any hard pomeron
contribution is no more than 10\% at $\surd s=1800$~GeV; because it falls
so rapidly as $\surd s$ is decreased, it becomes negligibly small
at HERA energies, even in hard or semihard processes.

In our calculations, we have had to decide how to choose the arguments
of the couplings $\alpha _s$. The choices we made were those that
enabled us to calculate most easily. In order to investigate how our
choices influence the output, we consider the simpler integral (3.12)
in which the energy-conservation constraint is removed.
We investigate 
its high-energy behaviour. (We should have preferred to discuss the
constrained cross-section (3.11), but we have found this to be too
difficult.) The asymptotic behaviour of (3.12) is controlled by 
the behaviour 
of the integrand for small $b$. So we need the small-$b$ behaviour of
$I(b,0)$, which comes from the small-$z$ region of the integration in
(4.2), that is $z$ less than some fixed $z_0$. In this region, we may replace
the Bessel function with unity. Then simple integration gives
$$
I(b,0)\sim C\log\left ({\log(z_0/b\Lambda)\over\log(\mu /\Lambda)}\right )
\eqno(7.1)
$$
and so the asymptotic behaviour of the contribution from values of 
$|{\bf b}|$ less than some fixed $b_0$ in the integral (4.1) is
$$
{2\pi ^3\over 81}\hat s^{\,-\C}\left ({\pd\over\pd\log\hat s}\right )^2\left\{
[\log (\mu ^2/\Lambda ^2)]^{-\C\log\hat s}\int _0^{b_0^2} db^2\;
[\log (z_0^2/b^2\Lambda ^2)]^{\C\log \hat s}\right\}~~~~~~~~~~$$$$
~~~~~~~~~~~~~~~~~~~~~~~~~~~~~~~~~~~~\sim \hbox{const } \hat s^{\C\log\log\hat s}
\eqno(7.2)
$$
This rise, faster than any power of $s$, is perhaps unexpected, though we know
of no basic reason to reject it. In any case, the inclusion of the energy-%
conservation constraint certainly slows the rise, and  ultimately, 
of course, it would be 
moderated by shadowing corrections. 
However, it is sensitive to how we make the
coupling $\alpha _s$ run, in particular how we play off the running 
$\alpha _s$ in the real-gluon BFKL insertions against that in the virtuals. 
As we have already mentioned, there is a delicate cancellation
between infrared real and virtual contributions, but the same is true of
the ultraviolet. The key point is that, although when we change from the usual
fixed $\alpha _s$ to a running one, both the real and virtual terms are 
suppressed, they enter with opposite signs.  Hence, by suppressing the virtuals
less than the reals we can actually increase the total output. In the
function $\phi (k^2)$ of (3.2a), which describes the  virtual
corrections, we chose to make $\alpha _s$ run with $k^2$, leading to
$\phi$ becoming constant at large $k^2$; see (3.7).
If we were to choose
instead to make $\alpha _s$ in the integration (3.2a) run with $q^2$, then
instead of $\phi (k^2)\to 2C$ for large $k^2$ we find $\phi (k^2)\to 2C\log\log
k^2$. Since each $k^2$ cannot be much larger than $s$, this is likely to
reflect itself in the fixed power $\hat s^{-\C}$ outside the integral (3.12a)
effectively 
being changed to $\hat s^{-C\log\log\hat s}$. This would then damp the 
$\hat s^{C\log\log\hat s}$ that comes from the integral, and perhaps 
leave something that in total looks much more like a fixed power. To decide
how to make $\alpha _s$ run in the different contributions would need a
nonleading calculation, which cannot be achieved at present, so the best
that we can say is that our calculation, which perhaps rises faster with
energy than it really should, already gives an output that is too small
to be relevant to experiment.

Finally, we note that
we have talked throughout about final-state ``partons'' rather than
``minijets'' because an observed minijet can achieve some of its high
transverse momentum by including fairly soft partons. We have not
attempted to separate this out from our calculations: it is buried within
the contributions we have called 
$\P _S\otimes\P _H$ etc --- see (1.2).
\vskip 10truemm
{\sl We thank Douglas Ross and others for stimulating criticisms.
This research is supported in part by the EU Programme ``Human Capital
and Mobility", Network ``Physics at High Energy Colliders'', contract
CHRX-CT93-0357 (DG 12 COMA), and by PPARC.}
\vfill\break\medskip\immediate\closeout\rfile\writestoppt
\baselineskip=8pt{{\bf References}}\medskip{\frenchspacing%
\parindent=20pt\escapechar=` \input refs.tmp\bigskip}\nonfrenchspacing
\bye